\documentclass[submission,copyright,creativecommons]{eptcs}
\providecommand{\event}{SCSS'2021} % Name of the event you are submitting to
\usepackage{underscore}           % Only needed if you use pdflatex.
\usepackage{etex}
 \usepackage{bussproofs} 
\usepackage[utf8]{inputenc} 
 \usepackage[T1]{fontenc}
\usepackage{fancyhdr}
\usepackage{rotating}
\usepackage{pdfsync}
\usepackage{enumerate}
\usepackage{graphics}
\usepackage{pdflscape}

\usepackage{amsmath,amssymb}
\usepackage{cite}
 
\usepackage{algorithmic}
 
\usepackage{algorithm}

\usepackage{bussproofs}

\title{\textsc{E-Cyclist}: Implementation of an Efficient Validation
  of FOL$_{\text{ID}}$ Cyclic Induction Reasoning}

\author{Sorin Stratulat\\
\institute{
Université de Lorraine, CNRS, LORIA\\
 Metz, F-57000, FRANCE\\
\email{sorin.stratulat@univ-lorraine.fr}
}
}

\def\titlerunning{\textsc{E-Cyclist}: Implementation of an Efficient Validation
  of FOL$_{\text{ID}}$ Cyclic Induction Reasoning}
\def\authorrunning{S.~Stratulat}
\begin{document}

\maketitle
\pagestyle{plain}

\begin{abstract}

 %\boldmath
  Checking the soundness of cyclic induction reasoning for
  first-order logic with inductive definitions
  (FOL$_{\text{ID}}$) is decidable but the standard checking
  method is  based on an exponential complement
  operation for Büchi automata. Recently, we introduced  a polynomial checking
  method whose  most expensive steps recall the comparisons done with
  multiset path orderings.  We describe the implementation of our method in the
  \textsc{Cyclist} prover. Referred to as \textsc{E-Cyclist}, it successfully checked all the proofs
  included in the original distribution of  \textsc{Cyclist}. Heuristics have been devised to
  automatically define, from the analysis of the proof derivations, the
  trace-based ordering measures that guarantee the soundness property.

  % FOL$_{\text{ID}}$ cyclic proof derivations may also be hard to
  % certify. \textsc{E-Cyclist} witnesses a strong relationship between the
  % two cyclic and well-founded induction reasonings. This opens the
  % perspective of using the known certification methods that work for
  % well-founded induction proofs.

\end{abstract}

%\pagenumbering{gobble}% Remove page numbers (and reset to 1)

%\thispagestyle{empty}

%------------------------------------------------------------------------- 
\paragraph{Introduction.}
% General induction principles: Noetherian Induction vs Descente Infinie

Cyclic pre-proofs for the classical first-order logic with inductive
predicates (FOL$_{\text{ID}}$) have been extensively
studied in~\cite{Brotherston:2005qy,Brotherston:2006uy,Brotherston:2011fk}. They
are finite sequent-based derivations
where some terminal nodes, called \emph{buds}, are labelled with sequents
already occurring in the derivation, called \emph{companions}. Bud-companion
(BC) relations, graphically represented as \emph{back-links}, are described
by an \emph{induction function} attached to the derivation, such that only
one companion is assigned to each bud, but a node can be the companion
of one or several buds. The pre-proofs can be viewed as digraphs whose
cycles, if any, are introduced by the BC-relations.

It is easy to build unsound pre-proofs, for example by creating a
BC-relation between the nodes labelled by the sequents from a
stuttering step. The classical soundness criterion is the \emph{global
  trace condition}. Firstly, the paths are annotated by traces built
from inductive atoms occurring on the lhs of the sequents in the path,
referred to as \emph{inductive antecedent atoms} (IAAs). Then, it is shown
that, for every infinite path $p$ in the cyclic derivation of a false
sequent, there is some trace following $p$ such that all successive
steps starting from some point are decreasing and certain steps
occurring infinitely often are strictly decreasing w.r.t. some
semantic ordering. We say that a \emph{progress point} occurs in the
trace when a step is strictly decreasing. A \emph{proof} is a
pre-proof if every infinite path has an infinitely progressing trace
starting from some point.

The standard checking method~\cite{Brotherston:2006uy} of the global
trace condition is decidable but based on an exponential complement
operation for Büchi automata~\cite{Michel:1988aa}. It has been
implemented in the \textsc{Cyclist} prover~\cite{Brotherston:2012fk}
and experiments showed that the soundness checking can take up to 44\%
of the proof time. On the other hand, we presented
in~\cite{Stratulat:2017ac,Stratulat:2018ab} a less costly,
polynomial-time, checking method. The pre-proof to be checked is
firstly normalized into a digraph $\cal P$ consisting of a set of
derivation trees to which is attached an extended induction function.
The resulting digraph counts the companions among its roots, as well
as the root of the pre-proof to be checked. Also, all infinite paths
in the pre-proof, starting from some point, can be reconstructed by
concatenating root-bud paths ($rb$-paths) in $\cal P$. Finally, a sufficient
condition for ensuring the global trace condition is to show that
every $rb$-path from the strongly connected components (SCCs) of
$\cal P$ has a trace that satisfies some trace-based ordering
constraints. Therefore, in theory, if the soundness of some pre-proof
can be validated with the
new method, it can also be validated with the standard one.% , where the ordering
% constraints are similar to those used for certifying cyclic Noetherian
% induction proofs~\cite{Stratulat:2017aa}.

%\pagenumbering{arabic}% Arabic page numbers (and reset to 1)

\paragraph{Implementation.}
\label{sec:impl}

Our method has been integrated in the \textsc{Cyclist} release
labelled as CSL-LICS14, by \emph{replacing} the standard checking
method. The result was called \textsc{E-Cyclist}. \textsc{Cyclist}
builds the pre-proofs using a depth-first search strategy that aims at
closing open nodes as quickly as possible. Whenever a new cycle is
built, model-checking techniques provided by an external model checker
are called to validate it. If the validation result is negative, the
prover backtracks by trying to find another way to build new cycles.
Hence, the model checker may be called several times
during the construction of a pre-proof.

Here is how our method works. Firstly, the pre-proof is normalized to
a digraph $\cal P$. To
each root $r$ from $\cal P$,
the method attaches a measure ${\cal M}(r)$ consisting of a multiset of IAAs of
the sequent labelling $r$, denoted by $S(r)$. One of the challenges is to find the
good measure values that satisfy the trace-based ordering constraints. A
procedure for computing these values is given by Algorithm~\ref{alg:alg}.

%\vspace{-.5cm}
\begin{algorithm}
\caption{GenOrd($\cal P$): to each root $r$ of $\cal P$ is attached a measure ${\cal M}(r)$}
\label{alg:alg}
\begin{algorithmic}
\FORALL{root $r$}
\STATE ${\cal M}(r):= \emptyset$
\ENDFOR 
\FORALL{rb-path $r\rightarrow b$ from a non-singleton SCC}
\IF{ there is a trace between an IAA $A$ of $S(b)$ and an IAA $A'$ of
  $S(r)$} 
\STATE add A to ${\cal M}(rc)$ and $A'$ to ${\cal M}(r)$, where $rc$
is the companion of $b$
\ENDIF
\ENDFOR
\end{algorithmic}
\end{algorithm}
%\vspace{-.5cm}

At the beginning, the value attached to each root is the empty set. Then, for
each $rb$-path from a cycle, denoted by $r\rightarrow b$, and for
every trace along $r\rightarrow b$, leading some IAA of $S(r)$  to
another IAA of $S(b)$, we add the corresponding IAAs to the values
of $r$ and the
companion of $b$, respectively. Since the number of rb-paths is
finite, Algorithm~\ref{alg:alg} terminates.

Algorithm~\ref{alg:alg} may compute values that do not pass the
comparison test for some non-singleton SCCs that are validated by the
model checker. For this case, we considered an improvement consisting
of the incremental addition of IAAs from a root sequent that are not
yet in the value of the corresponding root $r$. Since the validating
orderings are trace-based variants of multiset extension orderings, such an
addition does not affect the comparison value along the rb-paths
starting from $r$. On the other hand, it may affect the comparison
tests for the rb-paths ending in the companions of $r$. This may
duplicate some IAAs from the values of the roots from the rb-paths
leading to these companions. The duplicated IAAs have to be processed
as any incrementally added IAA, and so on, until no changes are
performed.

Table~\ref{table:tab} illustrates some statistics about the proofs of
the conjectures considered in Table 1 from~\cite{Brotherston:2012fk},
using inductive predicates as $N$, $E$, $O$, and $Add$, referring to
the naturals, even and odd numbers, as well as the addition on
naturals. All inductive predicates but $p$ are defined in~\cite{Brotherston:2012fk}. 
The proofs have been checked with the standard as well as our method.
The IAAs are \emph{indexed} in \textsc{Cyclist} to facilitate the
construction of traces; the way they are indexed influences how the
pre-proofs are built. Different indexations for a same conjecture may
lead to different proofs (see the statistics for the second and third
conjectures). The column labelled ‘Time-E’ presents the proof time
measured in milliseconds by using our method. Similarly, the `Time'
column displays the proof time when using the standard method, while
‘SC\%’ shows the percentage of time taken to check the soundness by
the model checker. ‘Depth’ shows the depth of the proof, ‘Nodes’ the
number of nodes in the proof, and ‘Bckl.’ the number of back-links in
the proof. The last column gives the number of calls for pre-proof
validations. The proof runs have been performed on a MacBook Pro
featuring a 2,7 GHz Intel Core i7 processor and 16 GB of RAM. It can
be noticed that, by using our method, the execution time is reduced by
a factor going from 1.43 to 5.

%\vspace{-.5cm}

\begin{table}[h!]
 
 \begin{tabular}{p{6.2cm}ccccccc}
    Theorem & \textbf{Time-E }& Time  & SC\% & Depth & Nodes & Bckl. &
                                                                  Queries \\ \hline
    $O_1x\vdash N_2x$ & \textbf{2} &  7 & 61 & 2 & 9 & 1 & 3\\
    $E_1x \vee O_2x \vdash N_3x$ & \textbf{4} & 11 & 63 & 3 & 19 & 2 & 6 \\
    $E_1x \vee O_1x \vdash N_3x$ & \textbf{2} & 9 & 77 & 2 & 13 & 2 & 6 \\
    $N_1x \vdash O_2x \vee E_3x$ & \textbf{3} & 7 & 52 & 2 & 8 & 1 & 4 \\
    $N_1x \wedge N_2y\vdash Q_1(x,y)$ & \textbf{297} & 425 & 40 & 4 & 19 & 3 &
                                                                    665
    \\
    $N_1x \vdash Add_1(x,0,x)$ & \textbf{1} & 5 & 76 & 1 & 7 & 1 & 4 \\
    $N_1x \wedge N_2y \wedge Add_3(x,y,z) \vdash N_1z$ &  \textbf{8} & 14 & 38 &
                                                                      2 & 8 & 1 & 16 \\
   $N_1x \wedge N_2y \wedge Add_3(x,y,z) \vdash Add_1(x,sy,sz)$& \textbf{15} & 22
                             & 32 & 2 & 14 & 1 &  14 \\
    $N_1x \wedge N_2y \vdash R_1(x,y)$ & \textbf{266} & 484 & 48 & 4 &35
                                                             & 5 &
                                                                   759\\
$N_1x \wedge N_2y \vdash p_1(x,y)$ & \textbf{597} & ? & ? & 4 & 28 & 3 &
                                                                       2315
  \end{tabular}

 \caption{Statistics for proofs checked
   with the standard and our method. \label{table:tab}}

\end{table}

%\vspace{-1cm}

\begin{figure}[h]
\centering 
\includegraphics[width=1\textwidth]{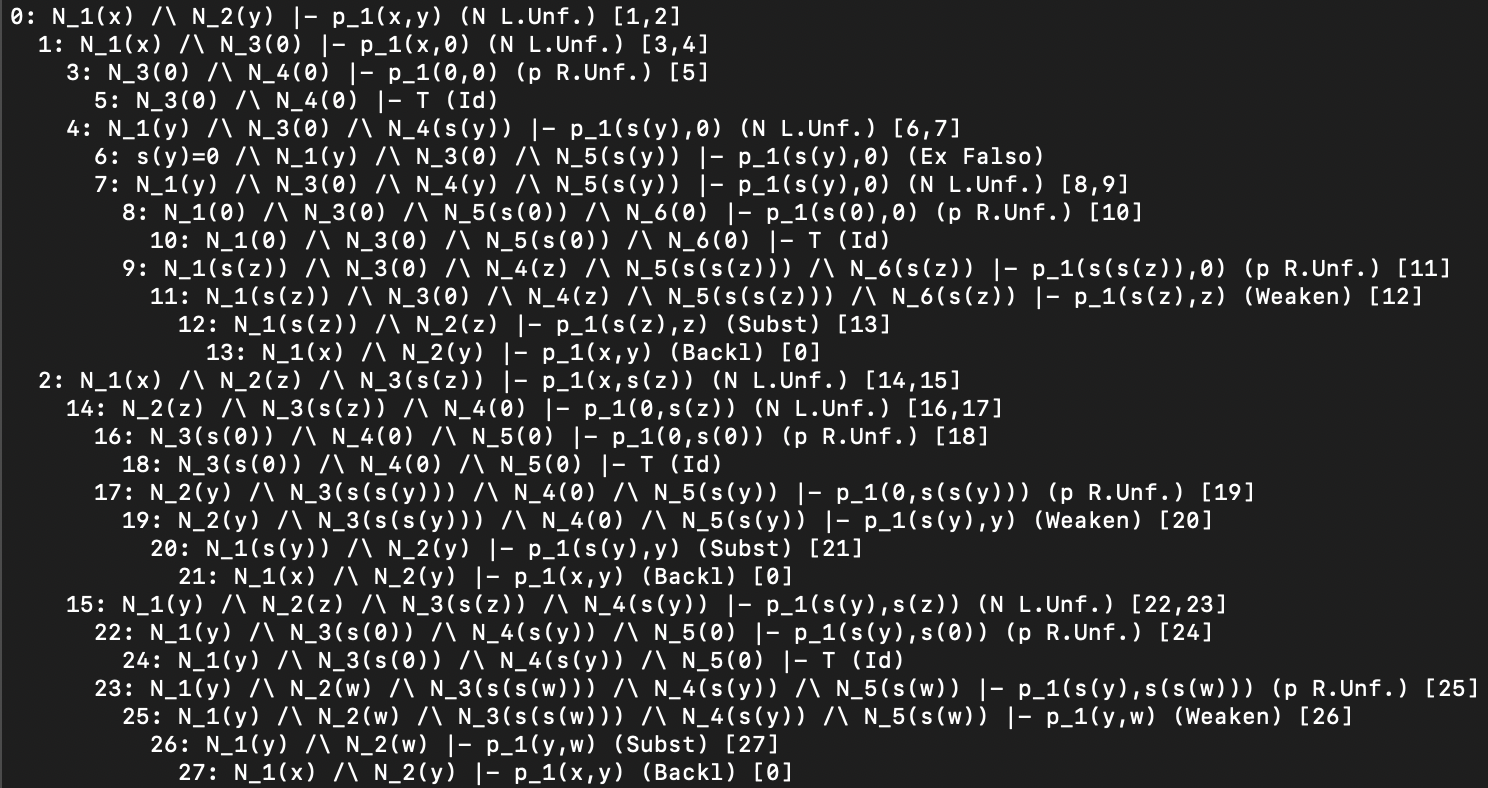}
\caption{The screenshot of the 2-Hydra pre-proof generated by \textsc{Cyclist}.}
\label{fig:hydra-cyclist}
\end{figure}

The last conjecture was not tested in~\cite{Brotherston:2012fk} and
refers to the 2-Hydra example~\cite{Berardi:2019aa}. A pre-proof of
it, reproduced in Figure~\ref{fig:hydra}, can also be generated by
\textsc{Cyclist}, as shown in Figure~\ref{fig:hydra-cyclist}. 
Unfortunately, \textsc{Cyclist} was not able to validate it using the
standard method, the missing figures being denoted by ?.

%\vspace{-.5cm}
\begin{figure}[!ht]
%\begin{scriptsize}
\begin{prooftree}
\def\ScoreOverhang{1pt}
\AxiomC{}
\UnaryInfC{$\vdash p00$}
\AxiomC{}
\UnaryInfC{$N0\vdash p10$}
\AxiomC{$(a) Nx,Ny\vdash pxy$}
\UnaryInfC{$Nsz,Nz\vdash pszz$}
\UnaryInfC{$Nsz,Nz\vdash pssz0$}
%\insertBetweenHyps{\hskip 2pt}
%\kernHyps{-1pt}
\BinaryInfC{$Nx'\vdash psx'0$}
\RightLabel{$Nx$}
\insertBetweenHyps{\hskip 3pt}
%\kernHyps{-1pt}
\BinaryInfC{$Nx\vdash px0$}
\AxiomC{}
\UnaryInfC{$N0,Nx\vdash px1$}
\AxiomC{$(a)Nx,Ny\vdash pxy$}
\UnaryInfC{$Nsu, Nu\vdash psuu$}
\UnaryInfC{$Nsu, Nu\vdash p0ssu$}
\AxiomC{$(a)Nx,Ny\vdash pxy$}
\UnaryInfC{$Nx',Nu\vdash px'u$}
\UnaryInfC{$Nsu,Nx',Nu\vdash psx'ssu$}
\RightLabel{$Nx$}
%\insertBetweenHyps{\hskip 2pt}
%\kernHyps{-1pt}
\BinaryInfC{$Nsu,Nx,Nu\vdash pxssu$}
\insertBetweenHyps{\hskip 3pt}
%\kernHyps{-1pt}
\BinaryInfC{$Nx,Ny'\vdash pxsy'$}
\RightLabel{$Ny$}
% \insertBetweenHyps{\hskip 2pt}
% \kernHyps{-1pt}
\BinaryInfC{$(a) Nx,Ny\vdash pxy$}
\end{prooftree}
%\end{scriptsize}
\caption{The Berardi and Tatsuta's cyclic pre-proof of the 2-Hydra example.}
\label{fig:hydra}
\end{figure}

% \begin{figure}[h]
% \centering
% \includegraphics[width=1\textwidth]{2-hydra}
% \end{figure}

\begin{figure}[h]
\centering 
\includegraphics[width=.65\textwidth]{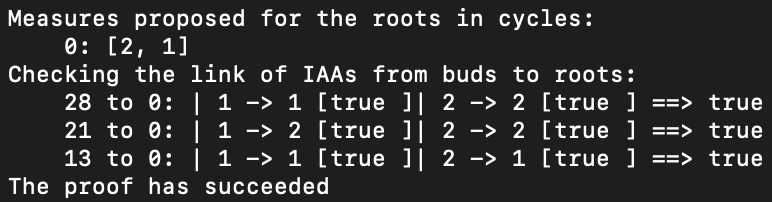}
\caption{The \textsc{E-Cyclist} validation of the 2-Hydra pre-proof from Figure~\ref{fig:hydra-cyclist}.}
\label{fig:hydra-proof}
\end{figure}

It also may occur that the proposed measure values, as shown in
Figure~\ref{fig:hydra-proof} for a non-optimised proof of 2-Hydra, may not pass some
comparison tests that succeed with the standard method, even when
using the improved version of Algorithm~\ref{alg:alg}. Indeed, this
happened while proving $N_1x \wedge N_2y \vdash R(x,y)$.
Luckily, the prover backtracked and finally found the same pre-proof as
that originally built with \textsc{Cyclist}.\footnote{The source code of the
implementation and the examples can be downloaded at
\url{https://members.loria.fr/SStratulat/files/e-cyclist.zip}}

We detail now how our method has been applied for
validating the 2-Hydra pre-proof from Figure~\ref{fig:hydra-cyclist}.

\paragraph{The 2-Hydra case.}
\label{sec:2-hydra}

\begin{figure}[!h]
\vspace{-2.5cm}
\begin{center}
    \includegraphics[width=\linewidth]{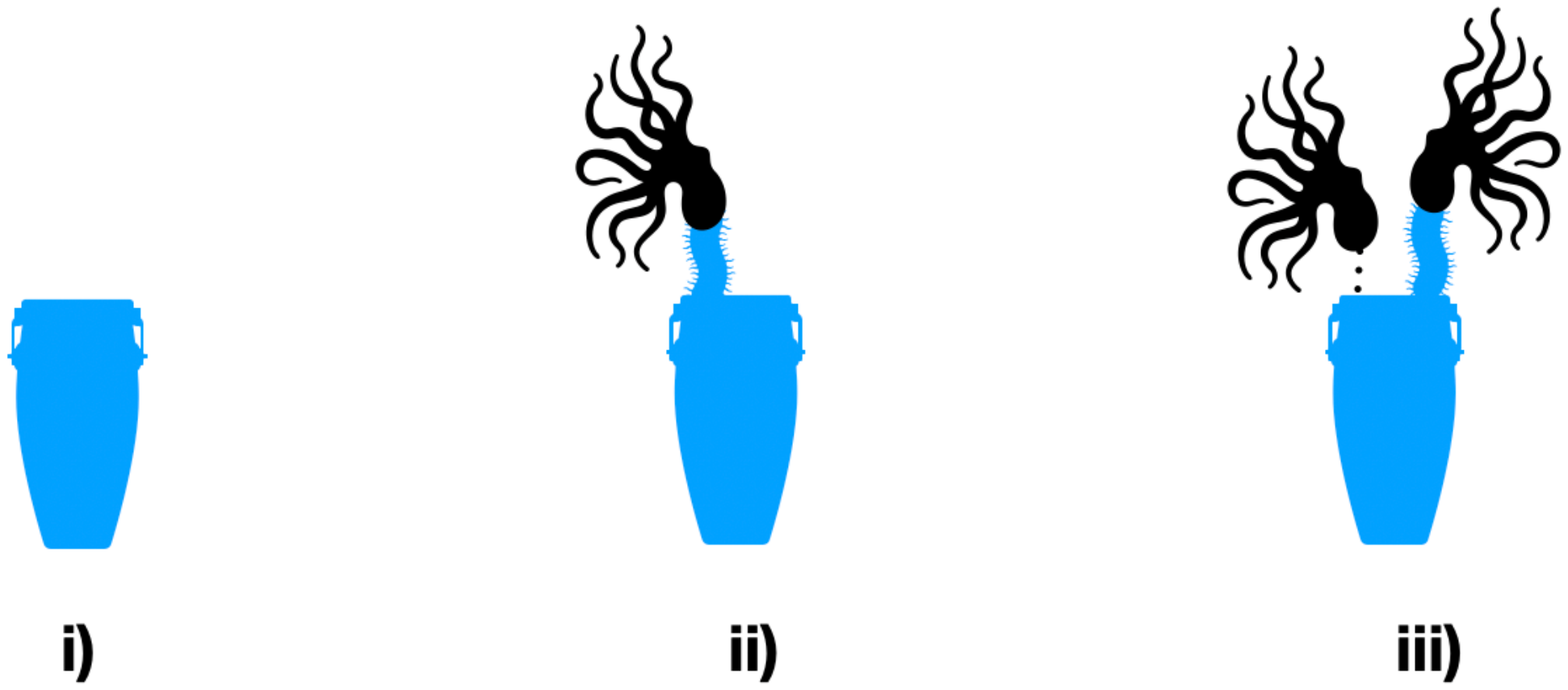}
  \end{center}
\vspace{-3cm}
\caption{\label{fig:hydra1} The cases when Hercules wins.}
\end{figure}

The 2-Hydra problem is a particular case showing the termination of
the battle between Hercules and Hydra~\cite{Dershowitz:2007bf} when
Hydra has at most two heads that hang on the top of necks of different
lengths. Hercules prevails if either Hydra has i) no heads at all, or ii) the length of
the first neck is 1 unit and it has lost the second head (i.e., the
length of the neck is 0), or iii) the length
of the second neck is 1 unit, as in Figure~\ref{fig:hydra1}.

\begin{figure}[!h]
\vspace{-2.5cm}
\begin{center}
    \includegraphics[width=\linewidth]{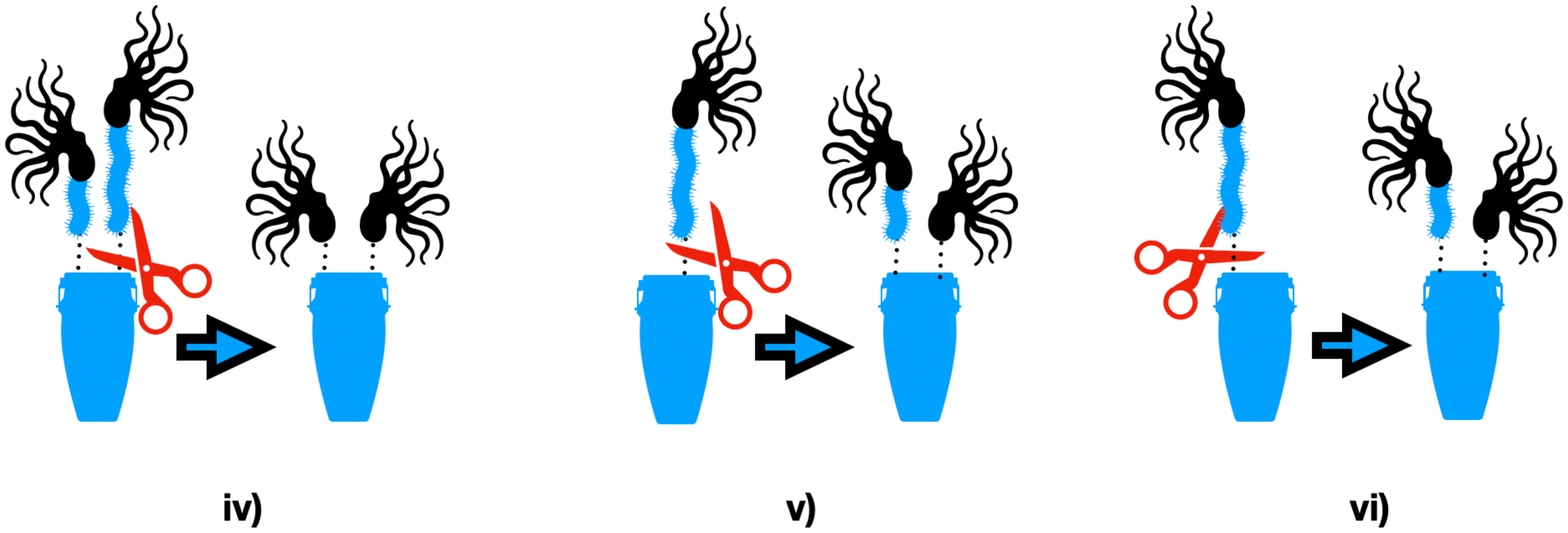}
  \end{center}
\vspace{-3cm}
\caption{\label{fig:hydra2} The cases when Hercules cuts the necks of Hydra.}
\end{figure}

Hercules can cut the Hydra's necks according to the following rules. If
both necks have strictly positive lengths, then Hercules can cut them such that
the first neck is shorter by 1 unit and the second by 2 units (see the
case iv in Figure~\ref{fig:hydra2}). If Hydra has already lost one of
the heads and the neck of the other head has a length $l$ of at least 2
units, the first head will have a neck of length $l-1$ units and the
second head a neck of length $l-2$ units (see the cases v and vi in
Figure~\ref{fig:hydra2}).

Next, we introduce the notations, the specification of the inductive
predicates, the inference rules, then explain the pre-proof from
Figure~\ref{fig:hydra-cyclist}. Contrary to the pre-proof from
Figure~\ref{fig:hydra}, the \textsc{Cyclist} pre-proof is horizontally
indented by the level of nodes. The nodes are numbered and labelled by
sequents where the comma (,) is replaced on the lhs of the sequents by
the conjunction connector ($/\backslash$). 

 The axioms defining the inductive predicates $N$ and $p$ are:

\begin{tabular}{lll}
\hspace{-1.9cm}\begin{minipage}[b]{.45\textwidth}
\begin{eqnarray}
\label{n0} \Rightarrow N(0) &\nonumber\\
\label{n1} N(x) \Rightarrow N(s(x))&\nonumber
          \end{eqnarray}
        \end{minipage} &
        \begin{minipage}[b]{.23\textwidth}
\noindent \begin{eqnarray}
\label{p0} \Rightarrow p(0,0)&\nonumber \\
   \label{p1} \Rightarrow p(s(0),0)& \nonumber\\
  \label{p2} ~~~~ \Rightarrow p(x, s(0))&\nonumber
\end{eqnarray}
\end{minipage}
&

        \begin{minipage}[b]{.45\textwidth}
\noindent \begin{eqnarray}
   \label{p3} ~~p(x,y)\Rightarrow p(s(x),s(s(y)))& \nonumber\\
  \label{p4} ~~~~p(s(y),y)\Rightarrow p(0,s(s(y)))& \nonumber\\
  \label{p5} ~~~~p(s(x),x)\Rightarrow p(s(s(x)),0)&\nonumber
\end{eqnarray}
\end{minipage}
\end{tabular}

\centerline{}
\centerline{}

\noindent The applied inference rule for each sequent is pointed out at the end of the sequent.

\textbf{($N$ L.Unf) [$n_1,n_2$]} generates the nodes $n_1$ and $n_2$
by choosing an IAA of the form $N(t)$. If $t$ is a variable, $t$ will
be replaced by $0$ and $s(z)$, where $z$ is a fresh variable. For the
second instantiation, the IAA is replaced by $N(z)$. This represents a
\emph{progres point}. If $t$ is of the form $s(t')$, the original
sequent is reduced to another sequent by replacing the chosen IAA
$N(s(t'))$ with $N(t')$.

\textbf{($p$ R.Unf) [$n$]} produces the node $n$ resulting from the
replacement of the consequent atom from the sequent labelling $n$ with the condition of
some axiom defining $p$ and whose conclusion matches the atom.

\textbf{(Id)} and \textbf{(Ex Falso)} delete trivial conjectures.
\textbf{(Weaken)} (resp., \textbf{(Subst)}) [$n$] is the LK's
weakening (resp., substitution) rule~\cite{Gentzen:1935fk} whose premise
labels $n$. Finally, \textbf{(Backl)} [$n$] shows that the
current node is a bud for the companion $n$.

The pre-proof from Figure~\ref{fig:hydra-cyclist} is already normalized
and has one non-singleton SCC with three rb-paths.
% The normalisation process ensures that, for every rb-path,
% \textbf{(Subst)} occurs only once and is applied just before  \textbf{(Backl)}.

% \begin{prooftree}
% \AxiomC{$\ldots$}
% \AxiomC{\ldots}
% \AxiomC{}
% \RightLabel{\textbf{(Backl)}}
% \UnaryInfC{\bf{bud}}
% \RightLabel{\textbf{(Subst)}}
% \UnaryInfC{\bf{$S'$}}
% \RightLabel{}
% \TrinaryInfC{\bf{$\vdots$}}
% \RightLabel{}
% \UnaryInfC{\bf{root} }
% \end{prooftree}

% \centerline{}

% \noindent where the only time when \textbf{(Subst)} is applied in the rb-path is
%  just before \textbf{(Backl)}.

Our validity method is based on properties to be satisfied
\emph{locally}, at the level of rb-paths. An rb-path $r \rightarrow b$
is \emph{valid} if $b$ is ``smaller'' than $r$ w.r.t. a trace-based
multiset extension relation. This relation guarantees the existence of
traces following each infinite path $p$, built from the
concatenation of the traces defined for the rb-paths along $p$. The
definitions for the standard and trace-based multiset extension are:

  \begin{itemize}
    \item (\emph{standard multiset extension}) $B <_{mul} A$ if there are
  two finite multisets $X$ and $Y$ such that $B = (A-X) \uplus Y$,
  $X\neq \emptyset$ and $\forall y\in Y$, $\exists x\in X, y < x$
  holds. 
  
\item  (\emph{trace-based multiset extension}) $b$ is ``smaller'' than $r$ if, after
  pairwisely deleting the IAAs linked by a non-progressing trace along
  $r \rightarrow b$ (the result is $X$ and $Y$ as above), $X\neq \emptyset$ and $\forall y\in Y$,
  $\exists x\in X$ such that there is a progressing trace along $r \rightarrow b$
  between $x$ and $y$.
\end{itemize}

In Figure~\ref{fig:hydra-proof}, we summarize
the result of the application of the improved version of Algorithm~\ref{alg:alg} to a non-optimized version of the pre-proof from
Figure~\ref{fig:hydra-cyclist}, for which the node 27 was denoted as
28. The found measure of the root is the multiset of its IAAs indexed
by 2 and 1, i.e., 
$\{N_2(x),N_1(y)\}$.

In Figure~\ref{fig:hydra-proof}, for each rb-path, $i$ -> $j$ denotes that there is a
trace linking the root IAA
indexed by $j$ to the bud IAA indexed by $i$, \texttt{[true ]} means
that the trace is progressing, and  `\texttt{===> true}' informs that the rb-path is
valid, as follows:

\begin{enumerate}
\item 0 to 28 (27 in Figure~\ref{fig:hydra-cyclist});  the possible traces following this path are: [$N_1(x),N_1(x),
  \underline{N_1(y)},N_1(y),\\N_1(y),N_1(y),N_1(x)$] and
  [$N_2(y),\underline{N_2(z)},N_2(z),\underline{N_2(w)},N_2(w),N_2(w),N_2(y)$],
\item 0 to 21;  the possible traces are: [$N_2(y),
  \underline{N_2(z)}, N_2(z), \underline{N_2(y)}, N_2(y), N_2(y),
  N_2(y)$] and $\\$[$N_2(y),\underline{N_2(z)}, N_2(z), N_5(s(y)), N_5(s(y)), N_1(s(y)), N_1(x)$],
  and
\item 0 to  13;  the possible traces are: [$N_1(x),N_1(x),\underline{N_1(y)},N_1(y),
  N_1(s(z)), N_1(s(z)), N_1(s(z)), N_1(x)$] and  [$N_1(x),N_1(x),N_4(s(y)),\underline{N_4(y)}, \underline{N_4(z)}, N_4(z), N_4(z),N_2(y)$].
\end{enumerate}

All the above traces are progressing, where the underlined
IAAs correspond to progress points. By definition, these rb-paths
are valid and conclude that the 2-Hydra pre-proof is a proof, by using
arguments as in~\cite{Stratulat:2017ac,Stratulat:2018ab}.

\paragraph{Conclusions and future work.} We have implemented in
\textsc{Cyclist} a more effective technique for validating
FOL$_{\text{ID}}$ cyclic pre-proofs which allows to speed up the proof
runs by 5. Besides its polynomial time complexity, an important factor
for its efficiency is the lack of the overhead time required to
communicate with external tools. In practice, our method can validate
pre-proofs that cannot be validated by the CSL-LICS14 release of
\textsc{Cyclist}. Even if we do not have yet a clear evidence, we
strongly believe that this also holds for the other way around, as
this might have happened for the $N_1x \wedge N_2y \vdash R(x,y)$ example.

The considered pre-proof examples are rather small. We intend to test
our method more extensively and on cyclic pre-proofs from domains other than
FOL$_{\text{ID}}$, e.g., separation logic. % We also plan to implement
% a certification tool for \textsc{Cyclist} proofs, similar to what has been
% developed for certifying (cyclic) well-founded induction proofs~\cite{Stratulat:2017aa} built with the
% SPIKE prover~\cite{Stratulat:2020ab}. 
% % This result allows to `interpret' FOL$_{\text{ID}}$ cyclic pre-proofs
% % in Coq~\cite{coq} as well-founded induction proofs. By using methods for
% % certification similar to those for formula-based Noetherian
% % induction reasoning, we pave the
% % way to a solution for certifying FOL$_{\text{ID}}$ cyclic reasoning,
% % in general, and \textsc{(E-)Cyclist} pre-proofs, in
% % particular.
% \footnote{For the reviewers, we detail in Appendix~\ref{sec:2-hydra}
%   how our method has been applied for validating the 2-Hydra
%   pre-proof. The full Coq specifications and proofs for certifying
%   the pre-proof from Figure~\ref{fig:hydra-cyclist} are given at 
% \url{https://members.loria.fr/SStratulat/files/hydra-coq.zip} }

\bibliographystyle{eptcs}

\newpage
%\bibliography{all}

\end{document}